# Characterisation of carbon dust produced in sputtering discharges and in the Tore Supra tokamak

C. Arnas[a*], C. Dominique[a], P. Roubin[a], C. Martin[a], C. Brosset[b], B. Pégourié[b]

[a] *Laboratoire PIIM, UMR 6633 CNRS-Universite de Provence , F-13397 Marseille, France*

[b]*Association Euratom-CEA, CEA Cadarache, CEA/DSM/DRFC,*

*F-13108 S$^t$ Paul Lez Durance, France*

**Abstract**

The sputtering of inside wall components of tokamaks can lead to the injection of supersaturated vapour in the edge plasmas. The resulting condensation favours the formation of clusters which can give rise to solid particulates by further accretion. Sputtering discharges are proposed to have highlight on the formation of carbonaceous dust observed in the tokamaks with graphite based wall components. The flux of the sputtered carbon atoms is evaluated in the conditions of our laboratory discharges as well as the evolution of their energy distribution. It is shown that a cooling mechanism occurs through collisions with the discharge argon atoms, leading to a nucleation phase. A comparison between the carbon structure of the resulting dust particles and a dust sample collected in the Tore Supra tokamak is proposed. The structural differences are discussed and can be correlated to specific plasma conditions.

*PACS:* 52.40.HF; 52.25.Vy; 61.46.+w

---

[*] Corresponding author. Tel.: 33 4 91 28 83 18; fax: 33 4 91 28 82 25

E-mail address: arnas@up.univ-mrs.fr

## 1. Introduction

In the tokamaks, plasma facing wall components (*PFCs*) are exposed to high particle fluxes which can induce high sputtering erosion. Off normal events like disruptions, arcing, instabilities can furthermore produce melting and/or evaporation of these *PFCs* [1]. Additional chemical erosion of the components in graphite based material releases hydrocarbon gases as methane, ethylene, acetylene [2]. A portion of the physical and chemical eroded material generates dust with a wide range of size and shape. In the micro-millimeter size range and more, one can observe small pieces and fragments coming from the walls [3], irregular grains coming from brittle redeposited layers, flat flakes from thin coatings and fibrous particles [3-6]. In the micro-nanometer scale, spherical primary particulates (*PPs*) are observed, piled up in macroscopic agglomerates in which the presence of nanotubes has also being evidenced [3]. The *PPs* can aggregate to form chains of particles and dense spheroids[7] as this occurs in plasma processing [8,9]. These *PPs* result either from the condensation of the vaporised material or from the multiple collisions between atoms and molecules released by the *PFCs.* In particular, radicals coming from hydrocarbon gases can polymerise to form macro-molecular precursors [10].

The ITER program has initiated different studies on the dust formation mechanisms, the produced quantities and their effects in fusion devices [11,12]. Among the used wall materials, the graphite has the largest capability to retain tritium. Carbonaceous dust is then considered as a safety hazard in the case of accidental device opening and a potential limit of the performances.

The results presented in this paper concern the carbonaceous particle formation from supersaturated carbon vapours coming from cathode sputtering in argon parallel electrode glow discharges. Conditions of supersaturation can be developped in tokamak plasma edge when arcs take place or when ions accelerated in sheaths, acquire energy higher than the sputtering threshold. In particular, this process is likely dominant in the Tore Supra (*TS*) tokamak [13], operating since 2000 with a toroidal wall component made in carbon fiber composite and so called: Toroidal Pumped Limiter (*TPL*).

The laboratory experimental conditions are presented in Section 2. In Section 3, a simple model is presented to estimate the carbon flux injected in these conditions from the cathode surface bombardment. Then, a numerical analysis showing the cooling of the sputtered atom during their collisions with argon atoms is proposed since it is assumed that the particulate precursors come from condensation. A better understanding of the dust formation can be provided by a correlation between the carbon structure in the atomic scale and the plasma characteristics. In Section 4, we propose structural analyses [9,14] of particles produced in our laboratory experiments and of a dust sample collected in January 2003 in a shadowed area of the *TPL* of *TS,* i.e. not directly exposed to high fluxes of the magnetized plasma. The differences are discussed in Section 5, according to specific conditions and plasma parameters of each case.

**2. Experimental set up of sputtering discharges**

Argon DC glow discharges are produced between two parallel electrodes separated by 5 cm. The lower one of 10 cm diameter is made in polycrystalline graphite and used as cathode. The upper one of 12 cm diameter, made of stainless steel, is grounded and used as anode. They are set in a cylindrical stainless steel vessel of 40 cm long and 30 cm diameter, equipped with four windows for optical diagnostics. Argon is admitted through a wall inlet and pumped at the bottom of the device by an oil diffusion pump, connected with a rotary pump. The residual pressure is lower than $10^{-6}$ mbar after several discharges. The experiments are performed at the argon pressure $P_{Ar} \sim 0.6$ mbar, without pumping flow. Rather low discharge currents of $I_0 \sim 70$ mA are used, resulting in discharge voltages of $V_K \sim -600$ V and providing current densities of ~ 9 A/m². These conditions allow carbon injection in the plasmas through cathode sputtering by ions and charge exchange (CX) neutrals.

The stability of the discharges is controlled through the spectral line evolution of $C_2$ dimer, one of the molecular precursors of solid particles, formed here. In our conditions, the intensity of the $C_2$ Swan band head (516.5 nm) remains constant during ~ 20 min. We therefore, perform successive discharges of 10 min, separated by several minutes during which the pumping flow is restored. The mean cathode and anode temperature measurements by thermocouples are 120°C and 60°C, respectively. As a consequence, we assume that a gas temperature of ~ 100°C is reached during the discharges.

The presence of the atomic hydrogen line $H_\alpha$ in the successive plasmas is assigned to the water residual pressure and/or to the wall outgassing during the discharges. Oxygen or other impurities have not been observed by optical emission spectroscopy.

Nano-scaled particulates are collected on stainless steel foils, placed on the grounded anode surface. Their shape, size, microstructure and chemical composition are analyzed by ex-situ diagnostics: Scanning Electron Microscopy (SEM), X-ray Absorption Near Edge Spectroscopy (XANES) and InfraRed (IR) absorption spectroscopy.

## 3. Flux and energy distribution of sputtered carbon atoms

### *3.1 Flux of sputtered carbon atoms*

At the intermediate pressure of $P_{Ar}$ = 0.6 mbar, several assumptions can be made: i) the ionization within the cathode sheath is negligible, ii) the sputtering comes from the bombardment of argon ions accelerated in the cathode sheath and of CX neutrals for which $\lambda_{CX} < d_s$, $\lambda_{CX}$ being the mean free path of the charge transfer and $d_s$, the cathode fall length. In such conditions, one can have a raw estimation of the energy of the incident particles on the graphite surface, assuming a linear electric field profile $E$(x), along the direction perpendicular to the surface, up to the negative glow. After integration and if one take $E(d_s) = 0$, the potential expression is given by:

$$V(x) = V_K/d_S^2 \ x\,(2d_S - x) \quad (1)$$

The energy that ions acquire between to CX collisions is then:

$$E_i = eV_K/d_S^2\,\lambda_{CX}\,(2d_s - \lambda_{CX}) \quad (2)$$

In our conditions, the negative glow location can be estimated with a cathetometer, at about $d_s \sim 4$ mm from the cathode. In the case of intermediate pressure and voltage, $\lambda_{CX}$ is relatively independent of the incident energy and is about $\lambda_{CX} \sim 260$ µm. So that, (2) provides an ion energy of about 100 eV.

The carbon flux is given by:

$$\Phi_C = \gamma\,(n_n\,u_n + n_i\,u_i) \quad (3)$$

where $n_{n,i}$, $u_{n,i}$ are the atom, ion density and velocity, respectively. $\gamma$ is the sputtering yield of graphite by argon particles of 100 eV energy. It is of the order of 0.4 % [15]. According to the ion flux continuity across the sheath and to the fact that a raw estimation of 16 charge transfers along $d_s$ leads to 16 times more rapid neutrals of 100 eV than ions, the expression (3) can be rewritten as:

$$\Phi_C = 17\gamma\,n_i\,u_i = 17\gamma\frac{I_0}{eS} \quad (4)$$

where S is the cathode surface. Then, (4) gives $\Phi_C \sim 5\ 10^{18}\ m^{-2}s^{-1}$ allowing the nucleation of particulates in gas phase.

*3.2 Energy distribution of sputtered carbon atoms*

Sputtering is a physical process whereby atoms in a solid target are ejected when the target material is bombarded by energetic ions or atoms. This process is largely driven by momentum exchange between the projectiles and the target atoms. In Thompson's model [16], the sputtering is the result of collision cascades induced in the material volume by incident particles of energy in the range of 100-1000 eV. In such a case, the energy distribution of the sputtered atoms, in the direction perpendicular to the surface, in vacuum is:

$$F(E) = P\,\frac{1-\left[(E_b+E)/\Lambda E_1\right]^{1/2}}{E^2\,(1+E_b/E)^3} \qquad (5)$$

with:

$$P = \frac{N\pi a^2 \Lambda E_a \eta\, D\Phi_1}{16}, \quad \Lambda = \frac{4\,M_1 M_2}{(M_1+M_2)^2} \quad \text{and} \quad E_a = \frac{2\,E_r (Z_1 Z_2)^{7/6}(M_1+M_2)}{M_2}$$

where $E_1$ is the energy of the projectile, $\Lambda E_1$ the maximum recoil energy, $M_1$ and $M_2$ are the respective masses of the incident particles and the target atoms. N is the atom density and $a = a_o/(Z_1\ Z_2)^{1/6}$ is the screening radius of the inter-atomic potential, $a_o$ being Bohr's radius. $E_a$ is the value of $E_1$ giving the distance of closest approach in a head-on collision. $\eta$ is a numerical factor of 0.52 corresponding to a Coulombian interaction potential. D is the nearest neighbor distance in the target, $E_r$ is the Rydberg energy and $\Phi_1$ is the incident particle flux, perpendicular to the target.

Fig. 1(a) provides the normalized energy distribution function EDF, calculated with $\Phi_1 = \Phi_C/\gamma$ injected in F(E), for two incident energies: 100 eV (full line), in the range of our experimental conditions and also 300 eV (dotted line). The EDF increases linearly to a maximum near ½ $E_b$ where $E_b$ is the binding energy of the solid ($E_b$ = 7.4 eV in the graphite case). The width is rather large and roughly independent of the incident energy as well as the position of maximum.

In the presence of a gas, the sputtered atoms lose their energy through collisions with the gas atoms. This energy loss $E_F$ is given by:

$$E_F = (E - k_B T_G) \exp[n \ln(E_f / E_i)] + k_B T_G \qquad (6)$$

where E is the energy of the sputtered atoms at the target surface calculated with (5), $T_G$ is the gas temperature, $E_f/E_i$ is the ratio of energies after and before collisions and n is the number of collisions occurring in the gas, over the distance *d*. *n* is given by:

$$n = \sigma d P_G / k_B T_G \qquad (7)$$

where $P_G$ is the gas pressure and $\sigma$, the cross section assuming hard sphere collisions.

The evolution of $E_F$ during collisions can be calculated [17] taking into account all the possible energies of the gas atoms. $k_B T_G$ in (6) must be replaced by $E_G$, a given energy in the Maxwell distribution of the gas. Thus, for each $E_G$, the energy loss is calculated for a fixed value of the kinetic energy E, in Thompson distribution and weighted by the collision probability. This latter parameter is given by the product of F(E) by the Maxwell-Boltzman distribution at $E_G$. In this way, $E_F$ is computed over all possible collision combinations.

Fig. 1(b) shows the normalized energy distributions of the sputtered atoms calculated at $d$ = 3 mm from the cathode for $P_{Ar}$ = 0.6 mbar and two incident energies: 100 eV (full line) and 300 eV (dotted line). The maximum is now shifted towards lower values and the width of the distributions is reduced. The maximum is near 0.1 eV and is relatively independent on the sputtering particle energy.

At $d$ ~ 6 mm from the cathode, the EDF becomes Maxwellian, indicating a total thermalisation of the carbon atoms with argon (same temperature of 100°C). As a consequence of this thermalisation, a supersaturated carbon vapor may exist in the plasma, favoring the nucleation process by condensation [18].

## 4. Experimental results

*4.1 Dust sample examples*

The SEM image in Fig. 2(a) shows an example of two kinds of carbonaceous particulates produced in discharges of various duration and collected on the anode surface. Small *PPs* of 20-30 nm size, likely formed by condensation, are visible all over the surface. Bigger particulates are also present with a cauliflower like texture observed in the samples we have analyzed and especially evidenced by transmission electron microscopy. The particulate in the left side of ~ 250 nm size is clearly composed of well separated radial columns, having a porous texture. The two other ones, in the right side of size ~ 180 nm are composed of very close columns, separated by fissures. This peculiar shape, cauliflower like has already been attributed to an accretion process [19,20] i.e. a growing phase by collection of neutral or ionic species. We assume here that the column deposition could develop on surface inhomogeneities of the *PPs* and be amplified by a reduced surface mobility and a carbon sticking coefficient close to 1 on the one hand and by shadowing effects on the deposition dynamics process [21], on the other hand. The limitation of the surface diffusion is usually assigned to low kinetic energy of the incident species, characteristic of the intermediate and high pressure discharges. C-H bonds whose presence is revealed by our structural analysis can also produce diffusion barriers. In addition, the very low grain temperature similar to that of gas, in comparison to that of graphite melting one can contribute to the formation of an irregular texture [22].

The shape and size of dust sucked up by a commercial vacuum cleaner on a shadowed area of the *TPL* of *TS* have also been analyzed. SEM investigations in the size range of 0.05-200 μm show essentially the presence of flakes coming from non adherent deposited thin layers. In the majority of the cases, they are formed by columns of inhomogeneous spatial distribution. An example is given in Fig. 2(b). The proposed magnification shows columnar depositions, separated by deep and sometimes, broad fissures, forming bunches of various sizes and various shapes. It is assumed here that the carbon deposition comes essentially from the sputtering of *TPL*. Indeed, during operation, this *PFCs* is sputtered by energetic $D^+$ ions. Being accelerated in the *TPL* sheath, they gain energy of about 500 eV (sputtering yield of ~ 2%) leading to the injection of carbon impurities in the plasma edge. Carbon return to the *TPL* in the form of $C^{5+}$, for instance, produces self sputtering [15] at a yield of ~ 20%.

### *4.2. Dust carbon structure*

The atomic arrangement of deposited layers on substrates and on the surface of particulates formed in gas phase depends on the plasma parameters and on the discharge geometry. Generally speaking, when carbon is deposited by plasma processing, a great variety of crystalline and disordered structures can be obtained because of the multiple orbital combinations of the carbon element. In the $sp^3$ hybridisation for instance, the carbon atoms have four valence electrons producing tetrahedral $\sigma$ bonding (C-C) as in diamond and in alkane molecules. In the tri-coordinated $sp^2$ hybridisation, three electrons produce three $\sigma$ bonds in a plane and the fourth produces a $\pi$ bond (C=C), perpendicular to the plane. This hybridisation is characteristic of the graphite, composed of parallel planes of aromatic rings and of alkene molecules like ethylene. A mixture of $sp^2$ an $sp^3$ hybridisations provides a disordered structure, the so called amorphous carbon [23] where the number of aromatic rings can dominate ($sp^2$ dominating). When hydrogen is incorporated, an increase of the $sp^3$ state occurs inducing structural defects. The identification of these carbon hybridisations has been driven in dust examples produced in our sputtering discharges and in *TS*.

*4.3. X-ray absorption spectroscopy*

The chemical bonds and the surface atomic structure (3-5 nm depth) of dust synthesized in our laboratory conditions (*Lab. sample*) and collected on the *TPL* surface of *TS* (*TPL sample*) have been evidenced by X-ray absorption near edge spectroscopy (XANES). With this diagnostic, the core electrons absorbing the incident X-ray, probe the unoccupied energy bands $\pi^*$ and $\sigma^*$, above the Fermi level. This investigation gives information on the crystalline or amorphous nature of the analysed samples. The experiments have been performed in LURE (Super ACO storage ring, Orsay, France) on the SA72 beam line equipped with a high-energy toroidal grating monochromator providing a resolution of ~ 300 meV at the C-*K* edge, in the total electron yield detection mode.

A comparison of XANES spectra is proposed in Fig. 3(a). In the *TPL sample* spectrum the peak at 285 eV is the signature of the C 1s $\rightarrow \pi^*$ transition of the $sp^2$ graphite hybridisation. The spectral structures above 290 eV result from excitonic resonance (291.7 eV) and multiple scattering (MS) resonances within the carbon atoms, in the graphite structure. The peak at 288.5 eV could be assigned to C-H bonds likely due to the cyclohexane solvent used to stick the powder on the sample holder. This spectrum is therefore characteristic of randomly oriented graphite crystallites of a few nanometer length [24].

The characteristic resonance of unconjugated $sp^2$ $\pi^*$(C=C) bonds is observed in the *Lab. Sample* spectrum at 284.7 eV, slightly shifted to lower energies compared to graphite. This shift could be due to the incorporation of hydrogen and oxygen in the atomic structure. Indeed, the peaks at 286.3 eV and 288.2 eV are at the energy position of the $\pi^*$(C=O) and $\sigma^*$(C-H) transitions, respectively and the shoulder around 289 eV is attributed to the $\sigma^*$(C-OH) transition [25]. However, optical emission spectroscopy measurements performed during the dust formation have not shown oxygen peaks. The presence of this impurity is therefore assigned to the sample air exposure, resulting in oxidation reactions and in carbon-water interactions that also produce C-H bonds. The $H_\alpha$ peak observation in the plasma emission originating either from the residual pressure or from the wall outgassing also indicate that hydrogen is incorporated in the dust during its growth phase.

The continuum region above 290 eV is made of two broad transitions centered around 292 eV and 300 eV, assigned to the σ* resonance of $sp^3$ (C-C) and $sp^2$ (C=C) hybridisation, respectively. These spectral features show that the particulates formed in the sputtering discharges are made of $sp^3$ tetra-coordinated carbon (C-C), structural defects due to C-H bonds and dangling bonds which make the particles extremely reactive. They also contain $sp^2$ tri-coordinated atoms (C=C) that could result from C=C alkene and aromatic clusters of very short range (<1 nm) dispersed into the solid.

The desorption of the *Lab. Sample* impurities has been obtained by annealing at 300, 450 and 700°C. The same XANES analysis has been performed on the sample after each thermal treatment without air exposure. At 300°C, the impurity peaks decrease but are still visible as shown in Fig. 3.(b) in comparison with the origin spectrum (100°C). At 450°C, close to the *TPL* maximum average temperature reached before January 2003, the desorption is achieved. The new spectrum shows that new carbon arrangements occur, in particular with an increase of the $sp^2$ σ*(C=C) resonance, around 300 eV. New aromatic rings are formed and the absence of MS fine structure indicates that the sample is amorphous. The overall shape is now characteristic of an amorphous carbon structure in which the size of the formed aromatic ring groups is no more than 1 nm [24]. For higher temperature (700°C), no significant structural difference is observed compared to 450°C.

Fig. 3(c) shows that the new carbon structure of the *Lab. Sample* does not reproduce the one of *TPL sample* also annealed at 450°C. In this latter case, the peak at 288.5 eV assigned to C-H bonds of the cyclohexane solvent has disappeared. Despite this thermal treatment, the *TPL sample* initial structure remains and this suggests a formation temperature higher than 450°C, in good agreement with a structure of randomly oriented graphite crystallites, deduced from the initial spectrum in Fig. 3(a).

The structural analysis has been completed by IR absorption spectroscopy.

*4.4. Infrared absorption spectroscopy*

IR absorption spectroscopy is a standard analytical tool for the detection of C-H (C-D) bonds. This diagnostic provides the stretching (noted ν) and bending (noted δ) vibrational modes of the analyzed material containing molecules. Spectra were recorded using an ATR (Attenuated Total Reflexion) spectrometer (Nicolet, Avatar 370 FTIR) with a single reflexion and a resolution of 4 $cm^{-1}$. Fig. 4 and Fig. 5 display IR absorbance spectra, in the wavenumber range 3600-700 $cm^{-1}$. They have been smoothed and are presented with no correction of neither baseline, nor water and carbon dioxide spectra.

The *TPL sample* spectrum is shown in Fig. 4. It is rather flat, the noised bands around 1520 and 1690 $cm^{-1}$ being present in the baseline. The absorbance increase in the right side is assigned to the dust diffusion, in the considered wavenumber range. The overall shape of this spectrum shows that there was no incorporation of impurities like hydrogen (deuterium) during the growth phase, the aromatic cycles being not active by IR absorption spectroscopy.

*Lab. sample* spectra are shown in Fig. 5(a) and 5(b) before and after annealing at 450°C, respectively. The broad absorption band at ~ 3342 $cm^{-1}$ in Fig. 5(a) is due to -O-H stretching vibration of $H_2O$, trapped in the porous network of the cauliflower like dust, during air exposure. The reactivity with water is also observed with the bands at 1276 and 1045 $cm^{-1}$, due to the stretching mode ν(C-OH).

The spectral band between 3000 $cm^{-1}$ and 2850 $cm^{-1}$ is the signature of $sp^3$ aliphatic groups with the $-CH_3$ asymmetric stretching vibration mode at 2951 $cm^{-1}$ and the $-CH_2$ asymmetric and symmetric modes at 2924 $cm^{-1}$ and 2855 $cm^{-1}$, respectively. The corresponding $-CH_3$ asymmetric and symmetric deformation vibrations are observed at 1455 and 1380 $cm^{-1}$, respectively.

The band at 1702 $cm^{-1}$ comes from the vibration ν(C=O), produced by oxidation after the device opening. An intensity increase of all the above mentioned bands has been observed with the air exposure time, confirming the dust reactivity.

The most emergent band at 1600 $cm^{-1}$ is due to in-plane, aromatic C=C symmetric stretching mode. Indeed, the $sp^2$ (C=C) ring vibrations become visible i.e. IR allowed when a high level of non-symmetric substitution occurs in the aromatic structure of an irregular network. The corresponding deformation vibration is observed at 910 $cm^{-1}$, signature of anthracene [26]. The vibration of an out-of-plane aromatic C-H bending mode with one or two adjacent H, is observed [27] at 867 $cm^{-1}$.

The presence of alkene groups in the internal structure could be evidenced by the minor peak of the out of plane, bending vibration δ (C=C-H) at 982 $cm^{-1}$ .

To perform IR absorption spectroscopy after annealing at 450°C, the sample has been again air exposed during several hours producing the broad absorption band ν(O-H) at ~ 3333 $cm^{-1}$, in Fig. 5(b). The thermal treatment has strongly decreased the $sp^3$ aliphatic group. In particular, the $-CH_3$ asymmetric stretching vibration mode at 2951 $cm^{-1}$ has disappeared and only remain the asymmetric and symmetric stretching vibrations of $-CH_2$ group.

The carbon rearrangement is evidenced by the presence now of the aromatic C-H stretching mode around 3050 $cm^{-1}$, characteristic of polycyclic aromatic hydrocarbons. This can be correlated to new peaks at 886 $cm^{-1}$ and 822 $cm^{-1}$, corresponding to out-of-plane bending modes of anthracene and of pyrene [28], respectively. The bands centered at 1095 and 994 $cm^{-1}$ could provide the corresponding in-plane, aromatic δ(C-H) bending vibration of pyrene and anthracene, respectively.

The in-plane aromatic C=C stretching symmetric mode, around 1600 $cm^{-1}$ is still visible but slightly red shifted.

These features confirm XANES analysis results: the particulates formed in the sputtering discharges contain $sp^3$ tetra-coordinated carbon (C-C and C-H) and dangling bonds making them extremely reactive. They also contain $sp^2$ tri-coordinated atoms (C=C) resulting on aromatic clusters and on alkene groups. Their annealing at 450°C leads to new carbon arrangement when impurities are released: new $sp^2$ tri-coordinated groups are generated in the form of pyrene and anthracene. The obtained new spectrum remains different from that of *TPL sample.*

**5. Discussion and conclusion**

The differences in the shape and in the carbon structure of dust produced in *TS* and in sputtering discharges can be explained by differences in the plasma conditions that generate them and by differences in their transport, after their formation. In particular, *TPL* dust comes from a surface: 1) not exposed directly to the plasma, the ion bombardment producing $sp^3$ hybridisation, 2) with an average temperature of about 250°C, the heating treatment producing $sp^2$ hybridisation. However, this temperature is too low to explain the graphite crystallites they contain, well evidenced by X-ray absorption spectroscopy and by Raman spectroscopy not presented here [14]. This important result could show that the analyzed dust have been transported from hotter regions and/or because of their bad adherence, they can have been heated by off normal events like disruptions [29]. Confirmation can be given by the fact that 1) their annealing treatment at 450°C does not produce structural change and 2) they do not contain deuterium [30] or only small proportion not detected by the proposed diagnostics. This later result could be of major importance for the deuterium inventory of *Tore Supra*.

The sputtering discharge particulates are formed by $sp^3$ tetra-coordinated carbon (C-C and C-H) and dangling bonds. They also contain $sp^2$ tri-coordinated atoms (C=C) resulting on aromatic clusters and on alkene groups. These features of amorphous carbon are characteristic of particles that have grown by accretion of low energy species originating from a combination of plasma parameters. On the one hand, medium gas pressure and low discharge power produce low carbonaceous gas temperature as shown in Section 3. A reduced mobility of the carbon species on the dust surface is then expected, amplified by C-H bonds, producing diffusion barriers. On the other hand, Langmuir probe measurements show that the floating potential with respect to the plasma potential is about -10 V. This value, higher than the graphite binding energy provides a range order of the energy that ions acquire in the dust sheath [31,32]. Therefore, the dust surface bombardment, mainly due to the contribution of $Ar^+$ could also participate to the formation of structural defects like dangling bonds. All these features may explain the atomic scale arrangement. As expected with such a structure, external thermal treatments show that the released impurities lead to new carbon arrangements in the form of $sp^2$ hybridization, i.e. carbonisation occurs.

In the nanometer scale, a reduced surface mobility could also explain the observed columnar deposition on surface inhomogeneities of the *PPs*. These inhomogeneities may be amplified by a carbon sticking coefficient close to 1 on the one hand and by shadowing effects on the deposition dynamics process [21], on the other hand. The cauliflower texture has not been changed by annealing in the temperature range chosen here.

**Acknowledgements**

The authors are very grateful to Y. Ferro for constructive discussions on chemistry and on chemical physics and to T. Angot for useful discussions on the surface analysis. This work is partially supported by the EURATOM-CEA Association in the framework of a LRC (Laboratoire de Recherche Conventionné) CEA/DSM-Université de Provence/PIIM.

## References

[1] J. Winter, Plasma Phys. Control. Fusion 40 (1998) 1201.

[2] J. Roth, E. Vietzke and A. A. Haasz, Suppl. Nucl. Fusion 1 (1991) 63.

[3] Ph. Chappuis, E. Tsitrone, M. Mayne, X. Armand, H. Linke, H. Bolt, D. Petti, J.P. Sharpe, J. Nucl. Mater. 290-293 (2001) 245.

[4] W. J. Carmack, M. E. Engelhart, P. B. Hembree, K. A. Mc Carty and D. A. Petti, “DIII-D Dust Particulate Characterization”, External Report INEEL/EXT-97-00702 (November 1997).

[5] A.T. Peacock, P. Andrew, P. Cetier, J.P. Coad, G. Federici, F.H. Hurd, M.A. Pick,

C.H. Wu, J. Nucl. Mater. 266-269 (1999) 423.
[6] J. P. Sharpe, V. Rohde, The ASDEX-Upgrade Experiment Team, Akio Sagara, Hajime Suzuki, Akio Komori, Osamu Motojima and the LHD Experimental Group, J. Nucl. Mater. 313-316 (2003) 455.
[7] J. Winter and G. Gebauer, J. Nucl. Mater. 266-269 (1999) 228.
[8] A. Bouchoule and L. Boufendi, Plasma Sources Sci. Technol. 2 (1993) 204.
[9] C. Arnas, C. Dominique, P. Roubin, C. Martin, C. Laffon, Ph. Parent, C. Brosset, B. Pégourié, J. Nucl. Mat. 337-339 (2005) 69.
[10] Ch. Deschenaux, A. Affolter, D. Magni, Ch. Hollenstein, P. Fayet, J. Phys. D: Appl. Phys. 32 (1999) 1876.
[11] J. P. Sharpe, D. A. Petti, H. W. Bartels, Fusion Engineering and Design 63-64 (2002) 153.
[12] G. Federici, C. H. Skinner, J. N. Brooks, J. P. Coad, C. Grisolia, A. A. Haasz, A. Hassanein, V. Philipps, C. S. Pitcher, J. Roth, W. R. Wampler and D. G. Whyte, Nucl. Fusion 41 (2001) 1967.
[13] E. Dufour, C. Brosset, C. Desgranges, R. Reichle, C. Lowry, R. Mitteau, J. Bucalossi, J. Gunn, P. Monier-Garbet, B. Pégourié, E. Tsitrone, P. Roubin, C. Martin, C. Arnas, P. Chappuis, Y. Corre, R. Guirlet , J. Hogan, T. Loarer and P. Thomas, 32$^{th}$ EPS Plasma Physics Conference, 27 June-1 July 2005, Tarragona, Spain, P-5.002.
[14] P. Roubin, C. Martin, C. Arnas, Ph. Colomban, B. Pégourié, C. Brosset, J. Nucl. Mat. 337-339 (2005) 990.
[15] W. Eckstein, C. Garcia-Rosales, J. Roth and W. Ottenberger, “Sputtering data”, IPP 9/82, Max-Planck-Institut fur Plasmaphysik, Garching bei Munchen, Feb. 1993
[16] M. W. Thompson, Philos. Mag. 18 (1968) 377.
[17] K. Meyer, I. K. Schuller and C. M. Falco, J. Appl. Phys. 52 (1981) 5803.
[18] R. C. Flagan, M.M. Lunden, Mater. Sci. Eng. A 204 (1995) 113.
[19] B. Ganguly, A. Garscadden, J. Williams and P. Haaland, J. Vac. Sci. Technol. A 11 (1993) 1119.
[20] D. Samsonov and J. Goree, J. Vac. Sci. Technol. A 17 (1999) 2835.

[21] J. Hua Yao, C. Roland and H. Guo, Phys. Rev. A 45 (1992) 3903.

[22] J. A. Thornton, J. Vac. Sci. Technol. A 4 (1986) 3059.

[23] J. Robertson, Mat. Sci. Eng. R 37 (2002) 129.

[24] L. Fayette, B. Marcus, M. Mermoux, G. Tourillon, K. Laffon, Ph. Parent, F. Le Normand, Phys. Rev. B 57 (1998) 14123.

[25] J. Stöhr, "NEXAFS spectroscopy", Springer Series in Surface Sciences, (1996), Springer.

[26] S. Califano, J. Chem. Phys. 36 (1962) 903.

[27] A. Galvez, N. Herlin Boime, C. Reynaud, C. Clinard and J. N. Rouzaud, Carbon 40 (2002) 2775.

[28] S. Califano, J. Chem. Phys. 39 (1963) 1016.

[29] Y. Corre, C. Brosset, E. Dufour, D. Guilhem, C. Lowry, R. Mitteau, P. Monier-Garbet, B. Pégourié, E. Tsitrone and S. Vallet, 32$^{th}$ EPS Plasma Physics Conference, 27 June-1 July 2005, Tarragona, Spain, P-5.003.

[30] C. Brosset, H. Khodja and Tore Supra team, J. Nucl. Mat. 337-339 (2005) 664.

[31] J. Goree, Plasma Sources Sci. Technol. 3 (1994) 400.

[32] C. Arnas, M. Mikikian and F. Doveil, Phys. Rev. E 60 (1999) 7420.

**Figure captions:**

**Figure 1:** Normalized energy distribution function of the sputtered carbon atoms (EDF) in the direction perpendicular to the graphite surface, for two incident energies of the argon sputtering particles: 100 and 300 eV, (a) in vacuum and (b) at d = 3 mm from the cathode, computed taking into account all the possible collision combinations with the argon atoms at T = 100°C.

**Figure 2:** (a) Dust grains produced in gas phase from graphite sputtering. Presence of small primary particles and of grains of higher size formed by radial columnar deposition, (b) detail of a flake collected on a shadowed area of the Toroidal Pumped Limitor surface of the Tore Supra tokamak. The proposed magnification shows columns forming bunches of various shapes and sizes.

**Figure 3:** XANES C-K edge spectra of a dust sample collected on a shadowed area of the Toroidal Pumped Limitor of Tore Supra: *TS sample* and on glow discharges: *Lab. Sample* in (a), spectra of *Lab. Sample* before annealing and after annealing at 300, 450 and 700°C in (b) and in (c), spectra of *TS sample* and of *Lab. Sample* after annealing at 450°C.

**Figure 4:** Infrared absorbance spectrum of a dust sample collected on a shadowed area of the Toroidal Pumped Limitor of Tore Supra in the wavenumber range 3600-700 $cm^{-1}$.

**Figure 5:** Infrared absorbance spectra of a sputtering discharge dust sample, in the wavenumber range 3600-700 $cm^{-1}$, after air exposure in (a) and in (b), after annealing at 450°C and air exposure.

Figure 1

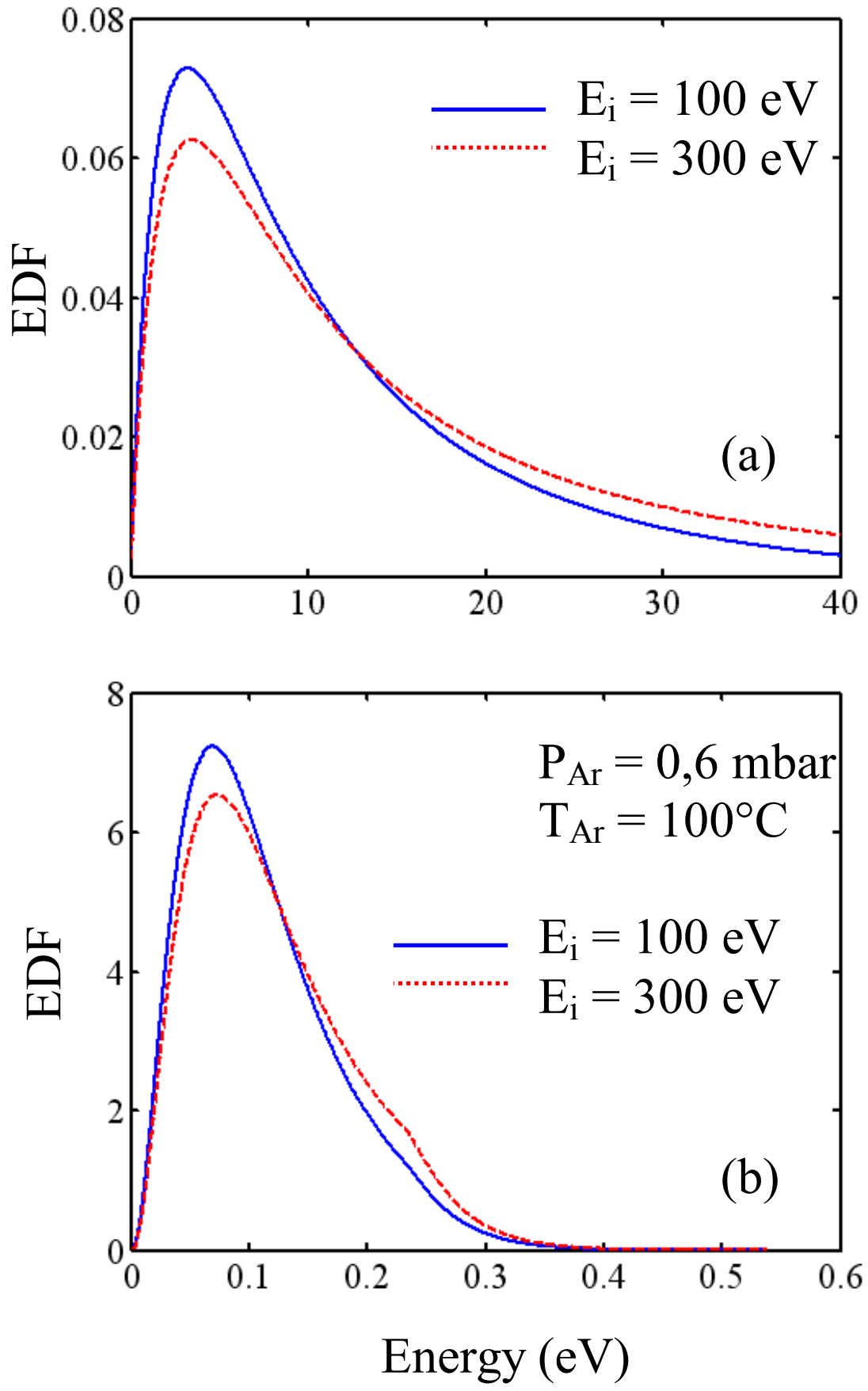

Figure 2:

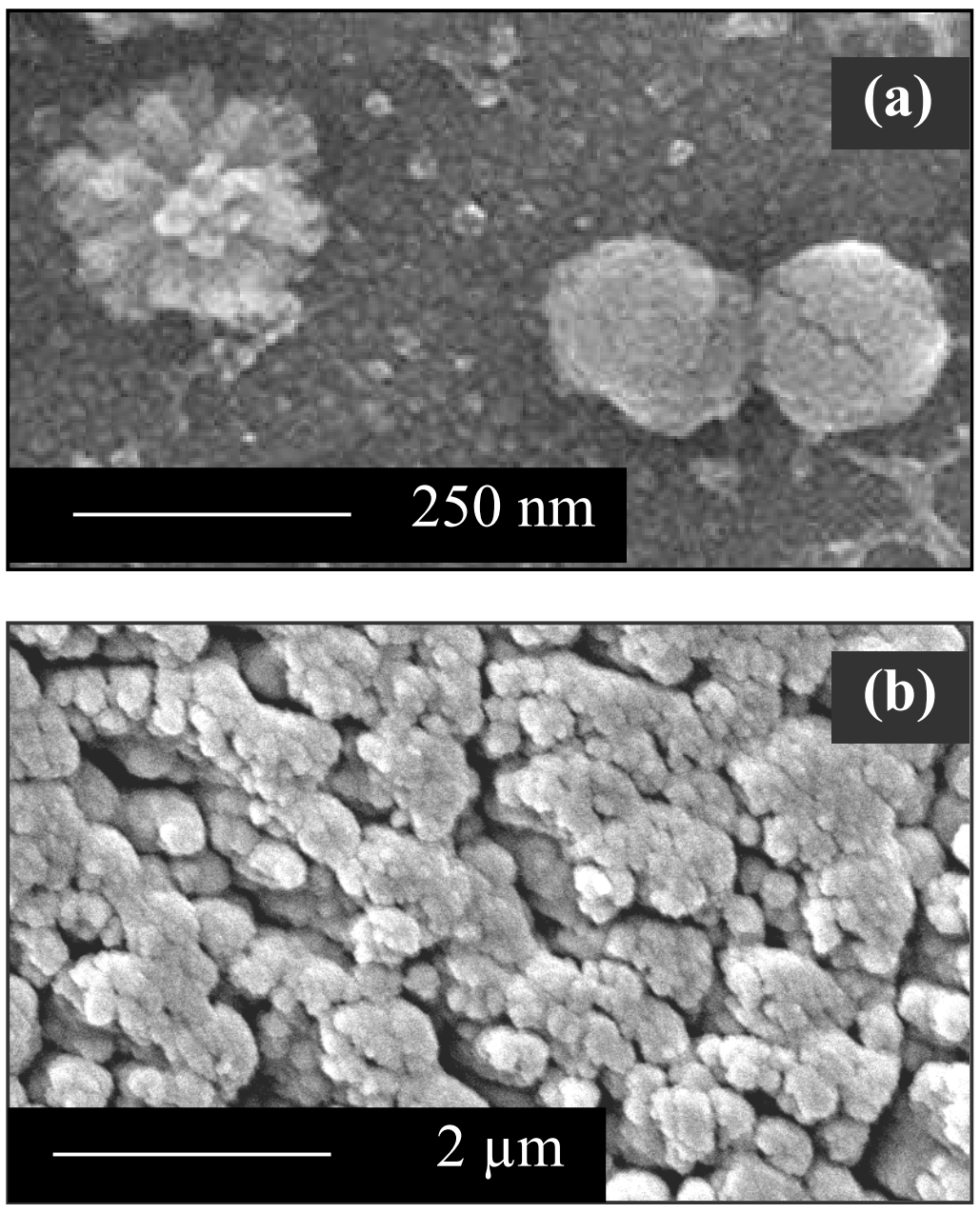

Figure 3:

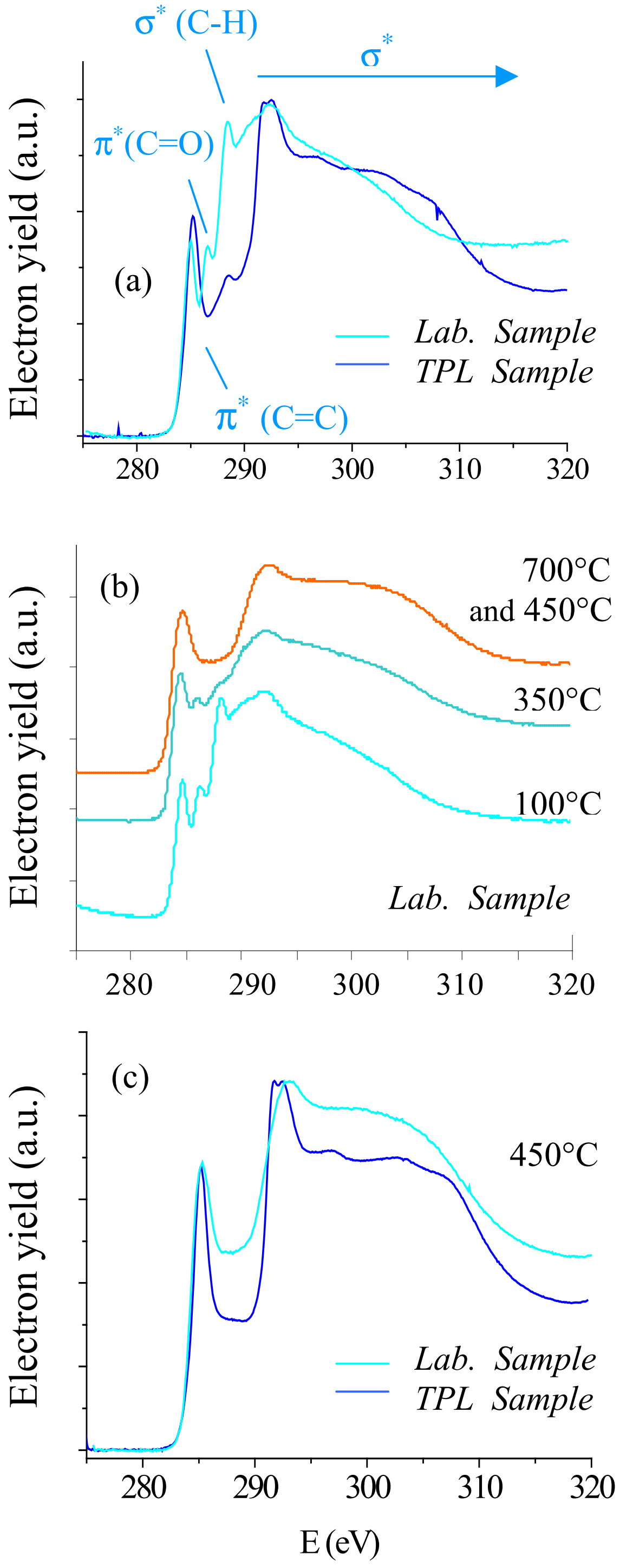

Figure 4:

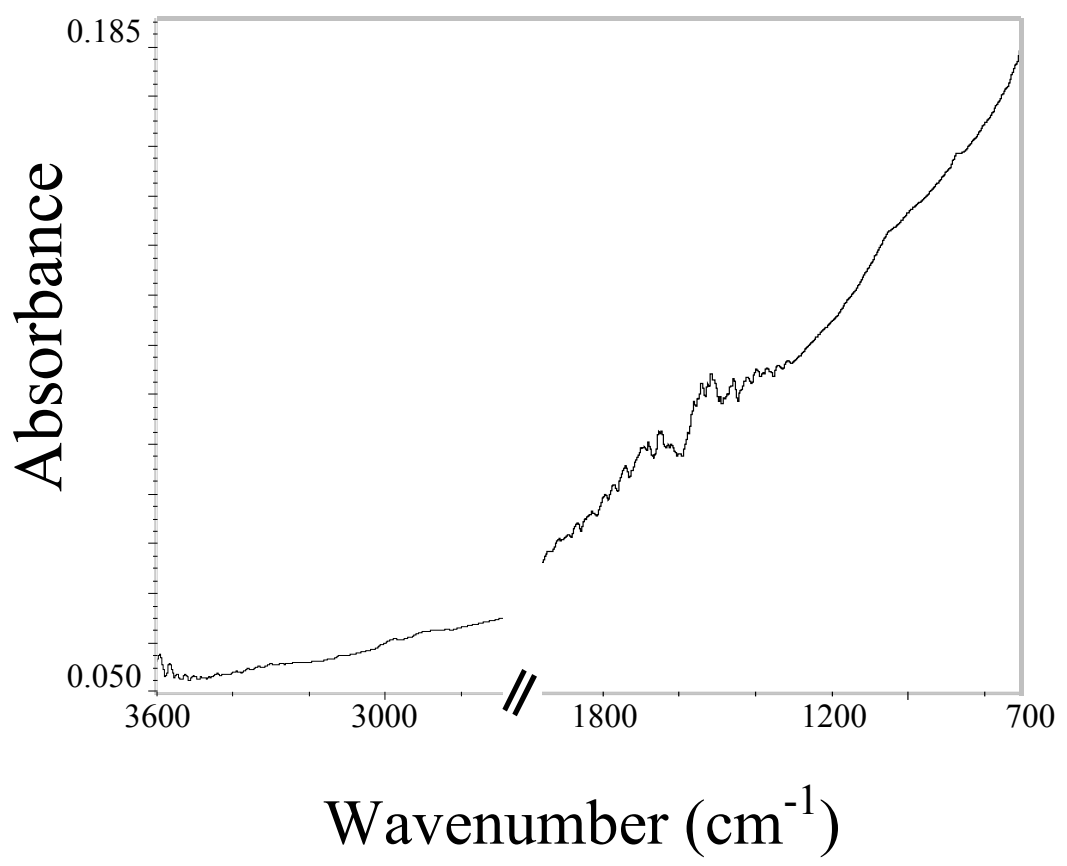

Figure 5:

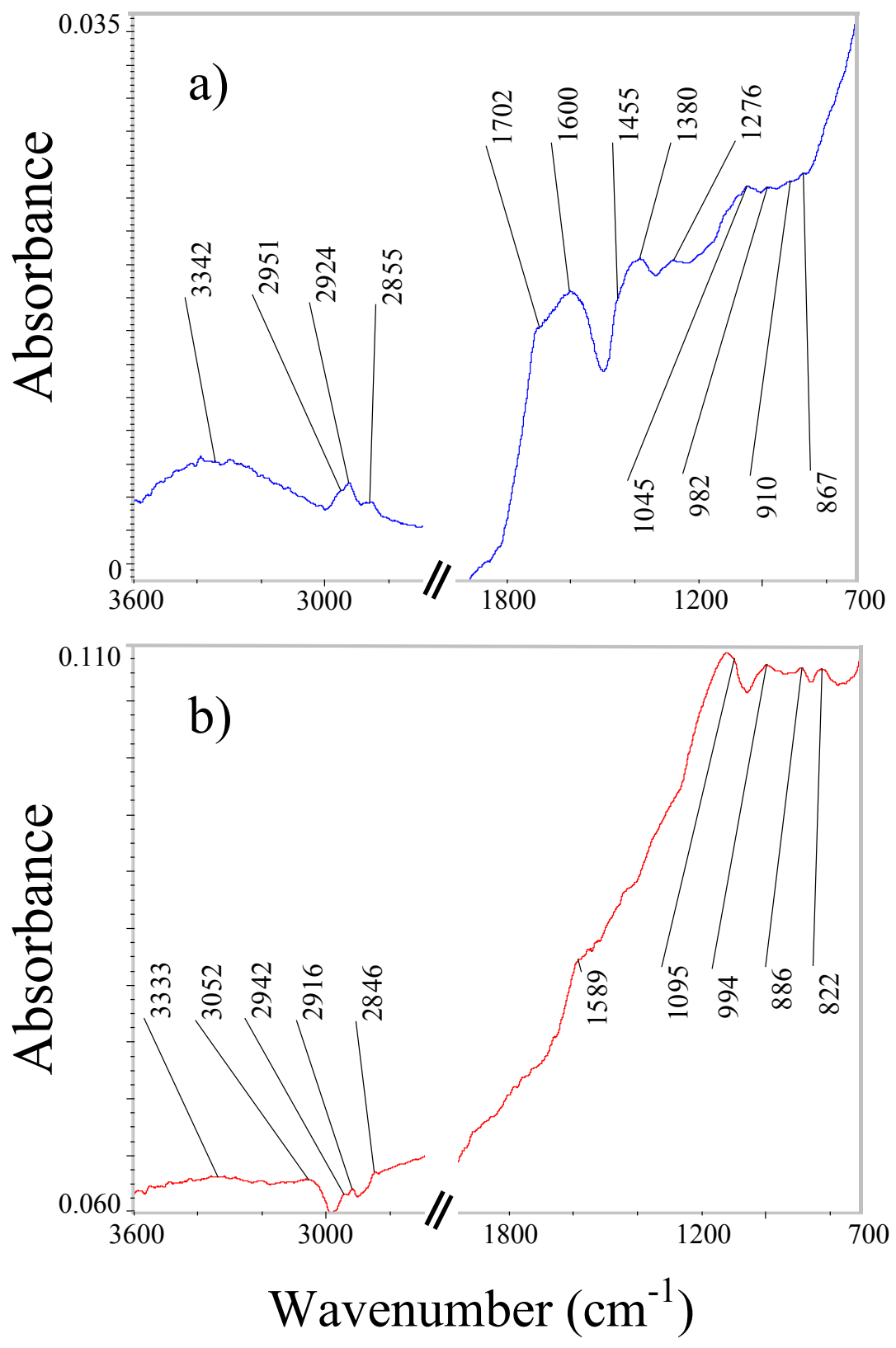